\documentstyle[psfig,aps,prl,amssymb]{revtex}

\begin{document}
\title{High-Temperature Expansions of Bures and Fisher 
Information Priors}
\author{Paul B. Slater}
\address{ISBER, University of California, Santa Barbara, CA 93106-2150\\
e-mail: slater@itp.ucsb.edu,FAX: (805) 893-7995}

\date{\today}

\draft

\maketitle

\vskip -0.1cm

\begin{abstract}
For certain infinite and finite-dimensional 
thermal systems, we obtain --- incorporating quantum-theoretic 
considerations into Bayesian thermostatistical investigations of
Lavenda --- high-temperature expansions over 
inverse temperature $\beta$ induced by volume elements (``quantum Jeffreys' 
priors) of Bures metrics. Similarly to Lavenda's results based on
 volume elements (Jeffreys' priors) 
of (classical) {\it Fisher information} metrics, we find
 that in the limit $\beta \rightarrow 0$ the quantum-theoretic 
priors either conform to Jeffreys' rule for variables over 
$[0,\infty]$, by being proportional to
 $1 / \beta$,
or to the Bayes-Laplace principle of insufficient reason, by being constant.
Whether a system adheres to one rule or to the other
 appears to depend upon
its number of
degrees of freedom. 
\end{abstract}

\pacs{PACS Numbers 05.30.-d, 42.50.Dv, 02.50.-r, 02.30.Mv}

\vspace{.1cm}

In this communication --- initially motivated by the investigation ``Bayesian
Approach to Thermostatistics" \cite{lav1} of Lavenda (cf.
 \cite{lav2,lav3,uff}) --- we examine certain prior distributions
 $\omega(\beta)$ over the inverse temperature
parameter $\beta$  that have recently been presented in the literature
\cite{slat0,slat1,para,kwek}. 
These distributions  are derived from ``quantum Jeffreys' priors'',
 that is, the 
volume elements $\mbox{d} V_{Bures}$ of the Bures/minimal monotone
metric \cite{hubner1,hubner2,caves,petzsudar}, for various
 finite and infinite-dimensional
convex sets of density matrices. We find that some, but not all, of these
derived priors
 satisfy --- in the high-temperature limit, 
$\beta \rightarrow 0$  --- Jeffreys' choice of
 ``$\omega(\beta) \propto 1/ \beta$, which is
invariant to transformations of the form $\zeta = \beta^{n}$, since 
$d \beta / \beta$ and $d {\beta}^{n} / {\beta}^{n}$ are always proportional.
This would not be true if the uniform distribution were used. Jeffreys cited
the measurement of the charge of an electron, where some methods give $e$
while others $e^{2}$, and certainly $\mbox{d}  e$ and $\mbox{d} e^{2}$ are not
proportional" \cite{lav1}. (Along these lines, let us emphasize 
for the purposes of this study  the obvious 
assertion that 
$\beta^{-1} \propto T$, where $T$ is the temperature.)

Lavenda \cite[sec. 4]{lav1} 
analyzed three models in particular: (1) an ideal monatomic gas having
the logarithm of its partition function 
$\propto  -{3 \over 2} \log{\beta}$;
(2) the harmonic oscillator with frequency $\nu$; and (3) a Fermi oscillator
with two levels. He determined that in the high-temperature limit the first
two of these yielded priors $\omega(\beta)$ proportional to $ 1/ \beta$,
while the third gave a constant prior.
Quite similarly to this set of  findings of Lavenda,
 all the prior distributions that we will examine
below are either proportional in the high-temperature limit to
$1 / \beta$ or to a constant. It is interesting to observe that
one of the infinite-dimensional systems we study --- the displaced thermal 
states \cite{para} --- has the same prior distribution, based on the quantum 
Jeffreys' prior, as that obtained for the Fermi oscillator 
by Lavenda\cite{lav1}, in his different analytical 
(classical) framework. Also, when we
attempt to apply the procedure of Lavenda to these states, as well as to the
displaced squeezed thermal states \cite{kwek,kwek0}, we find different high-temperature behavior
(that is, of the $1 / \beta$ type) 
than when we rely upon the quantum Jeffreys' priors. But, for the squeezed
thermal states \cite{twam}, the behavior using the two 
different (quantum and classical) approaches 
near $\beta$ is 
$1 / \beta$ in nature.

The term ``quantum Jeffrey prior'' was first employed in \cite{slat1}.
There, relying upon the innovative study  of Twamley \cite{twam} --- the
first to explicitly determine the Bures metric in an {\it infinite}-dimensional
setting --- a simple product (independence) form
\begin{equation} \label{twamvol}
\mbox{d} V_{Bures} = 
\upsilon(r) \omega( \beta ) \mbox{d} r \mbox{d}  \beta \mbox{d} \theta,
\end{equation}
was obtained
for the squeezed thermal states
\begin{equation}
\rho(\beta,r,\theta) =
   S(r,\theta) T(\beta) S^{\dagger}(r,\theta)/Z(\beta).
\end{equation} 
Here, $S(r,\theta)$ is the one-photon squeeze operator, and
\begin{equation} \label{twampart}
 Z(\beta) = (2 \sinh{\beta \over 4})^{-1}
\end{equation}
is a normalization factor (partition function)
chosen so that $\mbox{Tr}\rho =1$.
In the  form (\ref{twamvol}) (which we note
is independent of the unitary parameter
 $\theta$), 
 $\upsilon(r) = \sinh{2 r}$, and of more immediate
interest to the  investigation here,
\begin{equation} \label{twamvol3}
\omega(\beta) = { \cosh{\beta \over 4} \coth{\beta \over 4} \mbox{sech}{\beta \over 2} \over 8}.
\end{equation}
A series expansion in the vicinity of $\beta = 0$ yields
\begin{equation} \label{eqslat1}
\omega(\beta)= {1 \over 2 \beta} - {7 \beta \over 192} + {667 \beta^{3}
 \over 184320}
+ O[\beta]^{5}
\end{equation}
Near $\beta =0$, the first term predominates, so we discern that in the
high-temperature limit the 
$\beta$-dependent part (\ref{eqslat1}) of the quantum Jeffreys' prior
 (\ref{twamvol}),
in fact, satisfies the Jeffreys/Lavenda desideratum for a prior distribution
over $\beta \in [0,\infty]$ of being proportional to $1 / \beta$.

Subsequent to \cite{slat1}, Paraoanu and Scutaru \cite{para}
studied the case of the displaced thermal states.
They found the quantum Jeffreys' prior to be of the simple form
\begin{equation} \label{paravol}
\mbox{d} V_{Bures} =   {\mbox{sech} {\beta \over 2} \mbox{d} p
 \mbox{d} q \mbox{d}\beta \over  8},
\end{equation}
 where the variables $p$ and $q$ denote the 
displacements in momentum and position. Now, it is of interest to note,
that unlike (\ref{eqslat1}), this volume element can be normalized over the
full infinite range
$\beta \in [0,\infty]$ to a {\it proper}  prior probability distribution
function, $\omega(\beta) = {\mbox{sech}{\beta \over 2} \over \pi}$. (The mean of 
$\beta$ for this distribution is the product of
Catalan's constant, which is 
approximately 0.95966, and ${8 \over \pi}$, while the second
moment of $\beta$ is $\pi^{2}$.) Now, expanding around $\beta=0$, we have
\begin{equation} \label{eqpara}
\omega(\beta) = 
{\mbox{sech}{\beta \over 2} \over \pi} = {1 \over \pi} - {{\beta}^{2} \over 8 \pi}
+{5 {\beta}^4 \over 384 \pi} + O[\beta]^{6}.
\end{equation}
So, near $\beta = 0$, the prior behaves as a
uniform distribution, {\it not} fulfilling the Jeffreys/Lavenda desideratum.
(``This, however, is precisely the Bayes-Laplace rule, which Jeffreys considers
as an unacceptable representation of the ignorance concerning the value of
the parameter'' \cite{lav1}.)
The thermostatistical characteristics of this model for displaced thermal
states \cite{para} is essentially fully equivalent to those found by
Lavenda \cite{lav1} for a Fermi oscillator with
 two levels: 0 and $\epsilon_{0}$ (cf. \cite[eq. (3.5.11]{barnett}).
 (``As we have seen, the [Jeffreys']
invariance property also holds for Bose particles in the high-temperature
 limit.
However, the same is not true for Fermi particles'' \cite{lav1}. We have 
determined that this latter behavior also holds generically --- in the 
classical framework of Lavenda --- for  
the $SU_{q}(2)$ {\it fermion} model, relying upon its grand  
partition function \cite[eq. (23)]{ubriaco}.)

Kwek, Oh and Wang \cite{kwek} --- making use of the Baker-Campbell-Hausdorff
formula for quadratic operators \cite{wang1,wang2} --- then, extended
 these studies \cite{slat1,para}
to the displaced squeezed thermal states. They obtained the volume element
\cite[eq. (15)]{kwek},
\begin{equation} 
\mbox{d} V_{Bures}  = \lgroup {1 \over 2} \cosh^{2}{\beta \over 4}
\mbox{sech}^{3 \over 2}{\beta \over 2} \rgroup 
\sqrt{4 \cosh^{2} (2 r) -\sin^{2}(2 r)} \mbox{d} p \mbox{d} q \mbox{d} r
\mbox{d} \beta
\equiv \upsilon(r) \omega(\beta) \mbox{d} p \mbox{d} q \mbox{d} r \mbox{d} \beta.
\end{equation}
Now,
\begin{equation} \label{kwekvol}
 \omega(\beta) ={1 \over 2} \cosh^{2} {\beta \over 4} 
\mbox{sech}^{3 \over 2} {\beta \over 2}
= {1 \over 2} -{\beta^{2} \over 16} + {23 \beta^{4} \over 3072} +O[\beta]^{6}.
\end{equation}
So, similarly to (\ref{eqpara}) and unlike  (\ref{eqslat1}), this
univariate 
prior behaves {\it uniformly} in the immediate vicinity of $\beta=0$.
(The difference  between (\ref{eqslat1})
  and (\ref{kwekvol}), in this respect, is easily
evident in Fig. 2 of \cite{kwek}.)
Kwek, Oh and Wang noted that ``whereas the marginal probability distribution
for the undisplaced squeezed state diverges as $\beta \rightarrow 0$ or at
high temperature, in the case of the displaced squeezed state, the marginal
probability distribution goes to a finite value. The result is reminiscent of
a similar situation in chi-square distribution curves in which the probability
density function diverges at one degree of freedom, but not with higher
degrees of freedom. This analogy seems to indicate that the change in the
marginal probability density function in terms of inverse temperature stems
from an increased degree of freedom associated with the displacement of the
squeezed states'' \cite[p. 6617]{kwek}.
This line of argument suggests that 
perhaps a relation can be established between the degrees of
freedom of a system 
(in particular, the three instances analyzed above) 
and whether or not the associated prior $\omega(\beta)$
fulfills in the limit $\beta \rightarrow 0$ the Jeffreys/Lavenda desideratum 
or the Bayes-Laplace rule (or conceivably neither).

The three scenarios --- squeezed thermal states, displaced thermal 
states, and displaced squeezed thermal states --- examined above all
pertain to infinite-dimensional (continuous variable) quantum systems.
We now turn our attention to the cases of spin-${1 \over 2}$ and spin-1
(that is, two and three-level) systems.
Here, the quantum Jeffreys' priors, that is the volume elements of the
associated 
Bures metrics, are not typically parameterized in terms of  
inverse temperature parameters. 
So we can not immediately study the high-temperature limit but must have
recourse to a somewhat more indirect, but quite standard argument. That is,
we compute the one-dimensional (univariate) marginal distributions  of the 
(multivariate) quantum
Jeffreys' priors \cite{slat6}, which we interpret as densities-of-state or
structure functions, $\Omega(\epsilon)$.
Then, applying Boltzmann factors  and normalizing by the 
resulting partition 
functions $Z(\beta)$, we
 determine the corresponding canonical Gibbs distributions,
$\Omega^{*}(\epsilon | \beta) = \mbox{exp} [-\beta \epsilon -
\log{Z(\beta)}] \Omega (\epsilon)$. (We also note that Lavenda
 \cite[eq. (29a)]{lav1} considers, as well,  the 
different ``structure function''
$\Omega(\beta) = \omega(\beta) / Z(\beta)$, and the possibility of 
taking its Laplace transform to obtain a moment-generating function,
 $Y(\epsilon)$.) We use  the
contention of Lavenda \cite{lav1} (relying upon the asymptotic
equivalence between the maximum-likelihood estimate of $\beta$ and its
average value) that the implied prior 
(Bayes) distribution over $\beta$ should be taken to be
\begin{equation} \label{implied}
\omega(\beta) \propto \sqrt{var(\epsilon)} = \sqrt{ {\partial^{2} \over
\partial {\beta}^{2}} \log{Z(\beta})},
\end{equation}
where $var(\epsilon)$ is the variance of the energy --- that is,
$\langle (\epsilon - \langle \epsilon \rangle)^{2} \rangle$.
This is nothing other than the application 
 to the canonical distribution of the Bayesian/Jeffreys
procedure for constructing reparameterization-invariant priors. This consists
of taking the prior to be proportional to the volume element of the
(classical) Fisher information metric \cite{bernardo}.

For spin-${1 \over 2}$ systems, relying upon the Bures/minimal monotone
metric, one finds that \cite[eq. (12)]{slatbayes}
\begin{equation} \label{parthyper}
Z(\beta) = {2 I_{1} (\beta) \over \beta},
\end{equation}
where $I_{n}(\beta)$ denotes the modified 
(hyperbolic) Bessel function of the first kind.
Now, in this case,
\begin{equation}
\omega(\beta) = \sqrt{{\partial^{2} \log{Z(\beta})} \over \partial \beta^{2}} =
{1 \over 2} -{\beta^{2} \over 32} + {7 \beta^{4} \over 3072} + O[\beta]^{6},
\end{equation}
so, again, the Jeffreys/Lavenda desideratum of being proportional
to $1 / \beta$ is not satisfied, but rather the prior behaves uniformly 
in the vicinity of $\beta = 0$.
(``One method very common in statistical mechanics is the use of a 
high-temperature expansion: as $T \rightarrow \infty$ one tries to expand
the partition function as a series in powers of some parameter $\kappa (T)$ 
such that $\kappa(T) \rightarrow 0$ as $T \rightarrow
 \infty$'' \cite[p. 8]{lebellac}. In these studies, we expand not the partition
function {\it per se}, but the square root of the second derivative 
with respect to $\beta$  of its 
logarithm.)

Use of the {\it maximal} monotone metric (which is based on the 
{\it left} logarithmic derivative \cite{petzsudar}),
 in this case, rather than 
the Bures/minimal one (based on the {\it symmetric} logarithmic 
derivative), yields \cite[eq. (24)]{slat0}
\begin{equation}
Z(\beta) = ({\pi \over 2 \beta})^{1 \over 2} I_{1 \over 2} (\beta) =
{\sinh{\beta} \over \beta},
\end{equation}
leading to the arguably theoretically preferable Langevin function
\cite{tusz,lav4,langevin,yatsuya,brody},
\begin{equation}
{ \partial \log{Z(\beta)} \over \partial \beta} = -
\langle \epsilon \rangle = \mbox{coth} \beta - {1 \over \beta}.
\end{equation}
Nevertheless, the high-temperature behavior of the implied prior
 $\omega(\beta)$ --- again based on the relation 
(\ref{implied}) --- remains that of a constant near the origin, that is,
\begin{equation}
\omega(\beta) \propto {1 \over \sqrt{3}} - {\beta^2 \over 10 \sqrt{3}} +
{137 \beta^{4} \over 12600 \sqrt{3}} +O[\beta]^{6}.
\end{equation}

In \cite{slatjpa}, we studied certain {\it three}-level systems of the form
\begin{equation}
\rho = {1 \over 2} \pmatrix{v + z & 0 & x - i y \cr
0 & 2 - 2 v & 0 \cr
x + i y & 0 & v-z \cr},
\end{equation}
which are  one-parameter ($v$) extensions, in which the middle level
has become accessible, of the 
two-level systems,
\begin{equation} \label{nomiddle}
\rho = {1 \over 2} \pmatrix{ 1 + z & x - i y \cr
x + i y & 1-z \cr}.
\end{equation}
(We note that the full convex set of
 spin-1 density matrices is {\it eight}-dimensional in character 
\cite{bloore}.)
The univariate marginal probability distribution over $v$, 
obtained by integrating over the variables $x$, $y$ and $z$  in the normalized
quadrivariate Bures volume element 
\begin{equation}
p(v,x,y,z) ={3 \over 4 \pi^{2} v (1-v)^{1 \over 2} 
(v^{2} - x^{2} -y^{2} -z^{2})^{1 \over 2}}
\end{equation}
is \cite[eq. (19)]{slatjpa}
\begin{equation} \label{ext3}
\tilde{p} (v) = {3 v \over 4 \sqrt{1-v}}, \qquad 0 \leq v \leq 1.
\end{equation}
We interpreted (\ref{ext3}) as a density-of-states or structure function.
 We, then, determined \cite[eq. (42)]{slat0}
the associated partition function
\begin{equation}
Z(\beta) = {3 e^{-\beta} ((1+ 2 \beta) \sqrt{\pi} \mbox{erfi} (\sqrt{\beta})
-2 \sqrt{\beta} e^{\beta}) \over 8 \beta^{3 \over 2}}
\end{equation} 
(here $\mbox{ erfi}(z)$  represents the imaginary error
 function, that is  ${\mbox{erf} (iz) \over i}$)
by applying the Boltzmann 
factor $e^{-\beta v} = e^{- \beta \langle H \rangle}$ to (\ref{ext3}),
where
\begin{equation}
H = {1 \over 2} \pmatrix{1 & 0 & 0 \cr 0 & 0 & 0 \cr 0 & 0 & 1 \cr}.
\end{equation}
This leads --- {\it via} the argument of Lavenda \cite{lav1} again,
based on the relation (\ref{implied})  --- to
\begin{equation}
\omega(\beta) \propto {1 \over \beta} - {119 \beta \over 40}
 +{1891 \beta^{2} \over 140} + O[\beta]^{3},
\end{equation}
thus, satisfying the Jeffreys/Lavenda desideratum.
Since spin-${1 \over 2}$ particles are fermions and spin-1 particles are 
bosons, these results conform to Lavenda's assertion 
\cite{lav1} that priors associated
with bosons satisfy the Jeffreys' rule, while fermions do not.
We also note, somewhat in line with the discussion of Kwek, Oh and Wang \cite{kwek}, quoted above, that our spin-${1 \over 2}$
 example has an underlying three degrees of
freedom, while the spin-1 case has one more.

For the spin-${1 \over 2}$ systems (\ref{nomiddle}), the trivariate
(normalized) quantum Jeffreys' prior is
\begin{equation}
p(x,y,z) = {1 \over \pi^{2} (1-x^{2} -y^{2} -z^{2})^{1 \over 2}}.
\end{equation}
The univariate marginal probability distributions are of the form
\begin{equation} \label{univartwo}
 \tilde{p} (z) = {2 (1-z^{2})^{1 \over 2} \over \pi}
\end{equation}
Interpreting (\ref{univartwo}) as a density-of-states function, and 
using as the Hamiltonian,
\begin{equation}
H =  \pmatrix{1 & 0 \cr 0 & -1 \cr},
\end{equation}
one arrives
at the partition function (\ref{parthyper}).

Now, let us seek to apply the method for generating priors 
over $\beta$ of Lavenda 
directly to the three infinite-dimensional scenarios 
(squeezed thermal states, displaced thermal states and
displaced squeezed thermal states) first considered above, by taking for
the partition function $Z(\beta)$ in 
(\ref{implied}), the normalization factor that renders the
trace unity, so that one obtains a (properly normalized) density matrix.
For the squeezed thermal states, substituting (\ref{twampart}) into 
(\ref{implied}), we have
\begin{equation} \label{twamclass}
\omega(\beta) = {1 \over \beta} - {\beta \over 96} +{7 \beta^{2} \over
92160} +O[\beta]^{4},
\end{equation}
thus conforming to Jeffreys' rule --- under which $\log{\beta}$, not 
$\beta$ itself,  is distributed
uniformly.
For the other two types of infinite-dimensional thermal states considered
above, the result must be
the same as well, because $Z(\beta) $ takes the same form 
in them as (\ref{twampart}), since 
the displacement and squeeze operators are
unitary \cite[p. 4187]{wang2}. Contrastingly, based on the
 volume elements (quantum Jeffreys' priors) 
of the associated Bures metrics, as
we have noted in the first part of this letter, the prior (\ref{twamvol3})
 over $\beta$ for the squeezed thermal states does 
conform to the Jeffreys/Lavenda desideratum 
in the high-temperature limit, but the priors for the other
two, (\ref{paravol}) and (\ref{kwekvol}), follow the Bayes-Laplace 
principle of insufficient reason.
(One might then be puzzled by why, despite the unitarity of the squeeze 
and displacement operators, these three priors 
take different forms (cf. \cite{slaterhall}).)

It would, of course, be of interest to study 
invariance properties of prior distributions over the inverse 
temperature parameter $\beta$ for additional physical
scenarios, both in relation to  quantum Jeffreys' priors and the 
(Fisher information)
scheme of Lavenda for obtaining such distributions, and to elucidate
further any underlying governing principles.

 Let us note  the
assertion of Frieden and his associates that many physical laws 
have a Fisher information-theoretic basis \cite{frieden}.
In particular, Frieden, Plastino, Plastino and Soffer have ``shown that the
Legendre-transform structure of thermodynamics can be replicated without any
changes if one replaces the entropy $S$ by Fisher's information measure 
$I$''  \cite{friedenplastino}. Also, Grover's quantum search algorithm has
been demonstrated
 to be determined by a condition for minimizing Fisher information
\cite{alvarez}. In influential work,
Voiculescu \cite{voi} has developed analogues of the entropy and Fisher 
information measure for random variables in the context of 
{\it free} probability
theory. (Three different models of free probability theory are provided by
convolution operators on free groups, creation and annihilation operators
on the Fock space of Boltzmann statistics, and random matrices in the large
$N$-limit.)

In concluding, let us observe that Braunstein and Caves \cite{caves} derived
the Bures distance between two density operators
 by optimizing the Fisher information distance (obtained using the 
Cram\'er-Rao bound on the variance of 
any estimator) over arbitrary
generalized quantum measurements, not just ones described by one-dimensional
orthogonal projectors. Of course, the {\it volume elements} (Jeffreys' 
priors and quantum Jeffreys' priors) of the Fisher information and Bures 
metrics have been the basis for the thermostatistical investigation here.

\acknowledgments 

I would like to express appreciation to the Institute for Theoretical Physics
for computational support in this research and to K. \.Zyczkowski for his 
sustained interest in my work.

\end{document}